\documentclass{article}

\def\giorno{1/4/2004}

\def\a{\alpha}
\def\b{\beta}

\def\ga{\gamma}
\def\de{\delta}   
\def\eps{\varepsilon}
\def\phi{\varphi}
\def\la{\lambda}

\def\s{\sigma}

\def\th{\theta}

\def\h{{\cal H}}

\def\L{{\cal L}}

\def\P{{\cal P}}

\def\R{{\bf R}}

\def\Ga{\Gamma}

\def\La{\Lambda}
\def\Om{\Omega}

\def\pa{\partial}

\def\xb{{\bf x}}

\def\o+{\oplus}

\def\grad{\nabla}     
\def\ss{\subset}
\def\sse{\subseteq}

\def\<{\langle}
\def\>{\rangle}

\def\interno{\hskip 2pt \vbox{\hbox{\vbox to .18
truecm{\vfill\hbox to .25 truecm
{\hfill\hfill}\vfill}\vrule}\hrule}\hskip 2 pt}

\def\({\left(}
\def\){\right)}
\def\[{\left[}
\def\]{\right]}
\def\=#1{\bar #1}
\def\~#1{\widetilde #1}
\def\.#1{\dot #1}
\def\^#1{\widehat #1}
\def\"#1{\ddot #1}

\def\mapright#1{\smash{\mathop{\longrightarrow}\limits^{#1}}}

\def\EOE{ \hfill $\diamondsuit$ \medskip}

\begin{document}

\title{\bf Lie-Poincar\'e transformations and a reduction
criterion in Landau theory}

\author{Giuseppe Gaeta\footnote{e-mail: giuseppe.gaeta@mat.unimi.it ; g.gaeta@tiscali.it} \\
{\it Dipartimento di Matematica, Universit\`a di Milano} \\
{\it v. Saldini 50, I--20133 Milano (Italy)} }

\date{\giorno}

\maketitle

\noindent
{\bf Abstract.} {In the Landau theory of phase transitions one considers an effective potential $\Phi$ whose symmetry group $G$ and degree $d$ depend on the system under consideration; generally speaking, $\Phi$ is the most general $G$-invariant polynomial of degree $d$. When such a $\Phi$ turns out to be too complicate for a direct analysis, it is essential to be able to drop unessential terms, i.e. to apply a simplifying criterion. Criteria based on singularity theory exist and have a rigorous foundation, but are often very difficult to apply in practice. Here we consider a simplifying criterion (as stated by Gufan) and rigorously justify it on the basis of classical Lie-Poincar\'e theory as far as one deals with fixed values of the control parameter(s) in the Landau potential; when one considers a range of values, in particular near a phase transition, the criterion has to be accordingly partially modified, as we discuss. We consider some specific cases of group $G$ as examples, and study in detail the application to the Sergienko-Gufan-Urazhdin model for highly piezoelectric perovskites.}

\bigskip
\noindent {\tt PACS:} 05.70.Fh ; 64.60.-i ; 02.30.Oz \par\noindent
{\tt MSC:} 70K45; 70K50; 82B26 \par\noindent {\tt Keywords:}
Landau theory; phase transitions; normal forms; singularity
theory.

\vfill\eject

\section*{Introduction.}
\def\sn{0}

{\it Landau theory} is a powerful and widely used tool in analyzing second order phase transitions. This theory was introduced essentially on a phenomenological basis \cite{L1,Lan5}, but nowadays it can, in principles, be given a very rigorous justification in the framework of singularity theory \cite{ArnS}, or more precisely in that of catastrophe (or, in the russian parlance, perestroijka) theory \cite{ArnC,ArnB}.

In the same framework, one can also produce constructive techniques to simplify the power expansion describing the effective potential at the transition; thus {\it ``the art of throwing away inessential terms of Taylor series, retaining the higher order but physically important terms''} (quoted from \cite{ArnC}, sect. 5.5) can be, in principles, replaced by rigorous computations based on the so called ``spectral sequence'' \cite{ArnS,ToD}.

We have written twice, above, {\it ``in principles''}: indeed, the singularity theory approach to Landau theory is rigorous but extremely difficult to implement (beside being not so well known among the physicists who apply Landau theory to concrete problems), and would require quite involved computations even in the simplest case.

Thus, in practice, what happens is that once the relevant symmetry group $G$ has been identified (this is a physical -- and not mathematical -- matter, which we assume to be known by readers), and the general $G$-symmetric Taylor expansion for the effective potential $\Phi$ has been written down (in terms of $G$-invariant polynomials), one resorts to semi-empirical criteria to simplify the expression of $\Phi$.

The validity of these criteria is essentially proved {\it a posteriori} by checking that the simplified potential correctly describes the experimental results.

These simplifying criteria concern essentially two matters: $(a)$ the order $N$ at which the expansion can be truncated; $(b)$ the terms of order smaller than or equal to $N$ which are ``inessential'', i.e. that can be dropped without changing the qualitative properties of the potential $\Phi$.

As for point $(a)$, the essential criterion is that of {\it thermodynamical stability}: we require that there is some neighbourhood ${\cal D}$ of the origin which is invariant under the evolution determined by $\Phi$, i.e. such that $- \grad \Phi$ points inward at $\pa {\cal D}$. In practice, we require that the higher order terms have positive coefficients, so that at sufficiently large $|x|$ the potential $\Phi (x)$ is convex (see below for definition of the notation and a more precise statement).

In the present paper we want to discuss point $(b)$. A popular criterion for this \cite{Guf} is the following, which we quote almost verbatim from \cite{SGU} (the term ``minimal integrity basis'' appearing here will be explained below):

\medskip\noindent
{\bf Simplifying criterion.}
{\it Let $G$ be a compact Lie group, acting in $\R^n$ through a linear representation; let $\{ J_1 , ... , J_r \}$ be a minimal integrity basis for $G$, and let $F(J_1,...,J_r) : \R^n \to \R$ be a potential. Define, for $i=1,...,r$ and with $(.,.)$ the scalar product in $\R^n$, the quantities
$$ U_i (J_1,...,J_r) \ := \ \sum_{k=1}^r \, {\pa F \over \pa J_k} \ (\grad J_k , \grad J_i ) \ . \eqno(\sn.1) $$
A term can be omitted in the potential function without violating the type of extremal behaviour of this function if its coefficient is small enough and if it can be expressed as
$$ \sum_{i=1}^r \ Q_i (J_1 , ... , J_r) \ U_i (J_1 , ... , J_r ) \ + \ {\rm h.o.t.} \eqno(\sn.2) $$
with $Q_i$ polynomials in $J_1,...,J_r$ and ``h.o.t.'' denoting higher order terms.}
\medskip

In the present note we want to clarify the meaning of this criterion (see also \cite{SGU,Gufal}), which should be seen in the context of the orbit space approach to variational problems \cite{AbS,Sar1,Sar4,SarV,Mic2}, and discuss it in the light of the theory of (symmetric) Poincar\'e-Birkhoff normal forms \cite{ArnG,Elp,Wal,CGs,Gae02}. This will enable us to understand in simple terms the origin of the criterion, and how it should be modified when considering a range of values for the control parameter(s)

Thus, the {\bf goal of the paper} is twofold: $(i)$ on the one hand, provide a rigorous proof of a reduction criterion (see sect.3) closely related to the Gufan simplifying criterion quoted above when we consider the Landau potential at a given value of the control parameter(s); $(ii)$ on the other hand, show that when discussing a full range of values for the control parameter(s), we need some further restriction.

It should be stressed that Sanders \cite{San} has recently clarified the relation between normal forms, spectral sequence theory, and cohomological issues; however, the computations required in this framework are -- as for spectral sequences in general -- extremely involved already in the simplest case. We will stay at a simpler although less general level.

The {\bf plan of the paper} is as follows.
In section 1 we set our general notation and recall some general group-theoretical results to be used later on.
Section 2 is devoted to discussing the so-called Poincar\'e, and Lie-Poincar\'e, changes of coordinates.
In section 3 we apply these to a $G$-invariant potential; this will show that indeed a number of terms -- those satisfying a clear-cut condition -- can be eliminated by a well defined algorithm.
The discussion of sections 1-3 is conducted at a given value of the control parameter (which thus play little role in it) and provides mathematical justification for the reduction criterion, closely related to the simplifying criterion given above, stated at the end of  section 3.
In section 4 we discuss how this result applies to Landau theory, taking into due account the role of external parameters (such as temperature) controlling the phase transition. In section 5 we apply our discussion to a specific case, i.e. the Gufan-Sergienko model for highly piezoelectric perovskites \cite{SGU}; we will only discuss the form of the Landau potential to be considered, and not the physical consequences of it.

\medskip\noindent
{\bf Acknowledgements.} This work originated in discussions with
Yu. Gufan and I. Sergienko; I would like to thank them for these
and for communicating their work in advance. In the early stage of
this work I also had useful discussions with G. Pucacco and G.
Zanzotto, whom I warmly thank.

The financial support of {\it ``Fondazione CARIPLO per la Ricerca
Scientifica''} under the program {\it ``Teoria delle perturbazioni
per sistemi con simmetria''} (2000-2003) is gratefully
acknowledged.


\section{Basic notations and Landau theory}
\def\sn{1}

In this section we assume the reader has some general knowledge of Landau theory, as provided e.g. by \cite{Lan5} (see e.g. \cite{ToTo} for more detail), and -- as far as this is concerned -- just set the notation to be used below.

On the other hand, we need to recall some results concerning $G$-invariant polynomials, with $G$ a Lie group acting (linearly) in a finite dimensional space -- physically, the order parameter space; a huge mathematical literature is devoted to this subject. We will try to keep to as simple an exposition as possible; there are expositions designed for physicists rather than for mathematicians \cite{AbS,Sar1,Sar4,SarV,Mic2}, and the reader desiring more details is referred to these, see in particular the first part of \cite{AbS} and the first chapter of \cite{Mic2}. A readable introduction to the mathematical point of view is provided by \cite{DuK}.

\bigskip

We denote by $x \in M \simeq \R^m$ the order parameter, and by $G$ the group acting in $M$ to describe the symmetry of the system under study, or more precisely the symmetry inherited in the order parameters space.

We stress that $G$ acts through a (real) representation, i.e. a set of matrices $\{ T_g , g \in G \}$; however this is fixed once and for all, so that to avoid cumbersome notation we identify $g$ and $T_g$.

Due to a well known theorem of Palais and Mostow \cite{AbS}, we can always assume that the $G$ action is orthogonal after a suitable mapping to a higher dimensional space; however, the group actions met in applications of Landau theory are generally already orthogonal. We will thus assume that $T_g$ are orthogonal matrices; this implies that one of the $G$-invariants is always $|x|^2$.

The effective potential $\^\Phi (x) \in \R$ is a $G$-invariant polynomial, so that we should determine the most general polynomial in $x^1 , ... , x^m$ which is invariant under $G$.
The Landau potential $\Phi (x)$ will be a truncation of $\^\Phi$ to a suitable order $N$; moreover, we will be able to omit some ``unessential'' terms of order $n \le N$ (see below).

The polynomials $\^\Phi$ and $\Phi$ have coefficients depending on external parameters $\la$ (e.g., temperature, applied magnetic field,...); the state of the system is described by the minima of $\Phi_\la (x)$, which we denote as $x_\a (\la)$. Note that, in general, there will be different minima for a given value of $\la$; in particular, if $x (\la)$ is not a fixed point for the $G$-action, then the whole $G$ orbit through $x (\la)$ will be an orbit of minima
(if $G$ acts effectively in $M$, then by definition $0$ is the only fixed point for $G$). Moreover, there can be different $G$-orbits of minima for a given value of $\la$.

The symmetry of the state corresponding to $x (\la)$ will correspond to $G_{x(\la)}$, the isotropy group of $x (\la)$. We recall that by definition
$$ G_x \ := \ \{ g \in G \ : \ T_g x = x \, \} \ . $$
Note that if $x$ and $y$ are on the same $G$-orbit, then $y = T_g x$ for some $g \in G$, and then $G_y = g G_x g^{-1}$: there is a conjugacy class of isotropy subgroups associated to any $G$-orbit in $M$, and this is also called the {\it orbit type} $[Gx]$. A {\it phase} will be described precisely by an orbit type.

\medskip\noindent
{\bf Example 1.} Consider $G = {\bf Z}_2 = \{ e,g\}$ acting in $\R$ via $gx = - x$. Then $G_0 = G$, while for any $x \not= 0$, $G_x = \{ e \}$. An invariant Landau potential is of the form $\Phi (x) = - \la x^2 / 2 + x^4 /4$ (with $\la \in \R$), and as $\la$ changes sign we pass from a symmetric phase (orbit type $G$) to a non symmetric phase (orbit type $\{ e \}$). \EOE

\medskip\noindent
{\bf Example 2.} Consider $G = SO(n)$ acting in $\R^n$ through the defining representation. Then any point $x \not=0$ is left fixed by $SO(n-1)$ acting as rotations around the axis identified by the origin and $x$. All these subgroups are mapped one into the other by rotations in $SO(n)$. When we consider the potential $\Phi (x) = - \la x^2/2 + x^4 /4$ ($\la \in \R$) we observe a phase transition as $\la$ changes sign, and in this language it corresponds to passing from a phase with orbit type $[G\{0\}] = SO(n)$ to one of orbit type $[Gx] = SO(n-1)$. \EOE

In order to have a phase transition at a value $\la = \la_0$, it is needed that the orbit type $[G x(\la)]$ is not constant in a neighbourhood of $\la_0$, no matter how small.

In discussing what the ``suitable'' order $N$ is and which terms of order $n \le N$ can be omitted, we need to introduce some notions relative to $G$-invariant polynomials.

\subsection{$G$-invariant polynomials}

We will consider a compact Lie group $G$ acting linearly and orthogonally in the space $\R^n$ (many of the notions and results mentioned below have a much wider range of applicability, but here it suffices to consider this frame on the basis of physical needs).
We look at the ring of $G$-invariant scalar polynomials in $x^1,...,x^n$, denoted as $S(G)$.

By the {\bf Hilbert basis theorem} \cite{AbS,Hil,Olv}, there is a set $\{J_1 (x) , ... , J_r (x) \}$ of $G$-invariant homogeneous polynomials of degrees $\{d_1 , ... , d_r \}$ such that any $G$-invariant polynomial $\^\Phi (x)$ can be written as a polynomial in the $\{ J_1 , ... , J_r \}$, i.e.
$$ \^\Phi (x) \ = \ \^\pi \[ J_1 (x) , ... , J_r (x) \] \eqno(\sn.1) $$
with $\^\pi$ a polynomial in $(J_1,...,J_r)$.\footnote{It may be interesting to note that the theorem also holds for smooth functions \cite{Sch}.}

With our hypotheses on $G$, the algebra of $G$-invariant polynomials is finitely generated, i.e. we can choose $r$ finite. When the $J_a$ are chosen so that none of them can be written as a polynomial of the others\footnote{It is essential to understand that some of the $J_a$ could be written as non-polynomial functions of the others, and the
$J_\a$ could verify polynomial relations, see Example 5 below.} and $r$ has the smallest possible value (this value depends on $G$), we say that they are a {\it minimal integrity basis (MIB)}. In this case we say that the $\{J_a \}$ are a set of {\it basic invariants} for $G$.
There is obviously some arbitrarity in the choice of the $h$ in a MIB, but the degrees $\{ d_1 , ... , d_r \}$ of $\{J_1 , ... , J_r\}$ are fixed by $G$.\footnote{They are determined through the Poincar\'e series of the graded algebra $P_G$ of $G$-invariant polynomials, see \cite{Sar4}.}

We will from now on assume we have chosen a MIB, with elements $\{J_1 , ... , J_r\}$ of degrees $\{d_1 , ... , d_r \}$ in $x$, say with $d_1 \le d_2 \le ... \le d_r$.

\medskip\noindent
{\bf Example 3.} The simplest example is that of the fundamental representation of orthogonal groups themselves, $G = O(m)$, acting in $M=\R^m$. In this case (recall that invariants separate orbits) the only invariant function is $r = |\xb|$; however as we have to deal with polynomials in the coordinates $(x^1,...,x^n)$, the basic invariant will be $J_1 = r^2$. \EOE

\medskip\noindent
{\bf Example 4.} Consider $\R^2$ with coordinates $(x,y)$, and in this the group $G$ generated by the elements $g_x,g_y$ acting as $g_x : (x,y) \to (-x,y)$ and $g_y : (x,y) \to (x,-y)$. Basic invariants for this are obviously $J_1 = x^2$ and $J_2 = y^2$. \EOE

\medskip\noindent
{\bf Example 5.} Let us consider another simple example: $M = \R^2$, and $G$ is the group generated by simultaneous reflections in $x$ and in $y$, i.e. $G = \{ I , (R_x R_y) \}$ with $R_x : (x,y) \to (-x , y)$ and $R_y : (x,y) \to (x,-y)$. It is immediate to see that any $G$-invariant polynomial can be written as $\^\Phi(x,y) = \^\pi ( x^2 , y^2 , xy)$, i.e. we have a basis of three quadratic polynomials: $J_1 (x,y) = x^2 $, $J_2 (x,y) = y^2$, and $J_3 (x,y) = x y$. Note that here no elements of the MIB can be written as an algebraic (that is, polynomial) function of the others, but they are nevertheless satisfying an algebraic relation: indeed, $J_1 J_2 = J_3^2$. That is, they are {\it not} algebraically independent. \EOE

\medskip\noindent
{\bf Example 6.} Let us consider another simple example (relevant to the situation considered in \cite{SGU}): now $M= \R^3$, with $G$ the group generated by independent reflections in $x$, in $y$ and in $z$ (so, with the notation used in Example 5, $G$ is generated by $R_x$, $R_y$ and $R_z$). We have now a basis of three polynomials $J_\a (x,y,z)$, given by
$J_1 = x^2 + y^2 + z^2$, $J_2 = x^2 y^2 + y^2 z^2 + z^2 x^2$, $J_3  =  x^2 y^2 z^2$. \EOE

When the elements of a MIB for $G$ are algebraically independent, we say that the MIB is {\it regular}; if $G$ admits a regular MIB we say that $G$ is {\it coregular}. This case is, needless to say, easier to analyse; as the Example 5 above shows, the non-coregular cases can include very innocent-looking groups.

An algebraic relation between elements $J_\a$ of the MIB is said to be a relation (or, with a term of astronomical origin, a {\it sygyzy} \cite{Eva}) of the first kind. The algebraic relations among the $J$ are a set $S_{(1)}$ of polynomials in the $\{J_1 , ... , J_r \}$ which are identically zero when seen as polynomials in $x$; this admits a a minimal set of homogeneous generators $\{s_1^{(1)} , s_2^{(1)} , ... , s_{\s (1)}^{(1)} \}$. If there are algebraic relations among these, they are called relations (or syzygies) of the second kind, and so on.
A theorem by Hilbert guarantees that the chain of syzygies has finite maximal length (this is the homological dimension of the graded algebra $P_G$ mentioned above).

In the following we will need to consider a matrix built with the gradients of basic invariants, which we call (following Sartori \cite{Sar1,Sar4}) the $\P$-matrix. This is defined as
$$ \P_{ih} (x) \ := \ \< \grad J_i (x) , \grad J_h (x) \> \eqno(\sn.2) $$
with $\< .,.\>$ the standard scalar product in $M = \R^m$.

Note that the gradient of an invariant is necessarily a covariant quantity; the scalar product of two covariant quantities is an invariant one\footnote{Had we been considering a general manifold $M$ with metric $g_{ij}$, the scalar product between covariants quantities would be defined through $g^{ij}$, i.e. we would have $\P_{ih} (x) = (\grad J_i)_\a g^{\a \b} (\grad J_h)_\b$.}, and thus can be expressed again in terms of the basic invariants. Thus, {\it the $\P$-matrix can always be written in terms of the $J$ themselves}.

\medskip\noindent
{\bf Example 4} {\it (continued).} The basic invariants $J_1 = x^2$ and $J_2 = y^2$ are independent. Their gradients are $\grad J_1 = (2x,0)$, $\grad J_2 = (0,2y)$; the $\P$-matrix is hence
$$ \P \ = \ \pmatrix{ 4 x^2 & 0 \cr 0 & 4 y^2 \cr} \ = \ \pmatrix{ 4 J_1 & 0 \cr 0 & 4 J_2 \cr} \ . $$
\EOE

\medskip\noindent
{\bf Example 5} {\it (continued).} If $J_1 = x^2 \ge 0$, $J_2 = y^2 \ge 0$ and $J_3 = x y$, then as $(x,y)$ varies in $\R^2$, the point $(J_1 , J_2 , J_3 ) \in \R_+ \times \R_+ \times \R \ss \R^3$ varies on the manifold $J_3^2 = J_1 J_2$.

The gradients of the basic invariants are $ \grad J_1 = (2 x , 0 )$, $\grad J_2 = (0 , 2 y)$, $\grad J_3 = (y , x )$; note there is an obvious linear dependence among these at all points, which shows the $J$ are functionally dependent. The $\P$-matrix is
$$ \P (x,y) \ = \ \pmatrix{ 4 x^2 & 0 & 2 x y \cr 0
& 4 y^2 & 2 x y \cr 2 x y & 2 x y & x^2 + y^2 \cr} \ = \ \pmatrix{ 4 J_1 & 0 &
2 J_3 \cr 0 & 4 J_2 & 2 J_3 \cr 2 J_3 & 2 J_3 & J_1 + J_2 \cr} $$
\EOE

\medskip\noindent
{\bf Example 6} {\it (continued).} In this case we have
$$ \grad J_1 = 2 \pmatrix{x \cr y \cr z \cr} \ , \grad J_2 = 2 \pmatrix{(y^2 + z^2 ) x \cr (x^2 + z^2) y \cr (x^2 + y^2 ) z \cr} \ , \ \grad J_4 = 2 \pmatrix{(y^2 z^2) x \cr (x^2 z^2) y \cr (x^2 y^2 ) z \cr} \ . $$
Therefore, with straightforward explicit computations, the $\P$-matrix is
$$ \P  \ = \ 4 \ \pmatrix{J_1 & 2 J_2 & 3 J_3 \cr 2 J_2 & (J_1 J_2 + 3 J_3) & 2 J_1 J_3 \cr 3 J_3 & 2 J_1 J_3 & J_2 J_3 \cr} \ . $$
 \EOE

\subsection{Application to Landau theory}

Let us briefly comment on how the mathematical results mentioned above are relevant in Landau theory. As implied by the Hilbert basis theorem, any $G$-invariant polynomial, such as the effective potential $\^\Phi (x)$ or the Landau potential $\Phi (x)$, can be expressed as a polynomial $\^\pi (J(x))$.

In this way the evaluation of the map $\^\Phi : M \mapright{} \R$ is in principles substituted by evaluation of two maps, $h: M \mapright{} \Om$ and $\^\pi : \Om  \mapright{} \R$; here we have denoted by $\Om \sse \R^r$ the target space for ${\bf J} = (J_1 , ... , J_r)$.

However, if -- as is the case in Landau theory -- we have to consider the most general polynomial in $S(G)$, we only have to deal with the map $\pi : \Om \to \R$.

Note that for coregular groups we have simply $\Om = \R^r$, while for non-coregular groups $\Om$ is a submanifold of $\R^r$.

The space $\Om$ is also known as the {\bf orbit space} for the $G$ action on $M$; indeed, its points are in one-to-one correspondence with the $G$-orbits in $M$.

In facts $\Om$ is a semialgebraic manifold, i.e. a submanifold in $\R^r$ defined by algebraic equalities and inequalities; it is moreover a stratified manifold, i.e. the disjoint union of smooth manifolds of possibly different dimensions, with bordering relations related to the conjugacy class of isotropy subgroups on the $G$-orbits represented by points on each of these manifolds. We will not discuss these points, but just refer to \cite{AbS,Sar1,Sar4,SarV,Mic}.

As the $J$ are nontrivial homogeneous polynomials, the origin of $\R^m$ will correspond to $J_1 = ... = J_r = 0$, and a neighbourhood of zero in $M$ will be mapped to a neighbourhood of zero in $\Om$; thus we will not have to actually evaluate the map $J: M \to \Om$, but can just focus on $\^\pi : \Om \to \R$ in a neighbourhood of the origin.

We will boldly summarize our discussion as the following\footnote{Louis Michel (1923-1999) pioneered the use of orbit space techniques in Physics and Nonlinear Dynamics, originally motivated by the study of hadronic interactions; see e.g. \cite{Mic2,Mic,MicRad}. For an essential bibliography of his works on this theme, see \cite{ZhilMic}.}

\medskip\noindent
{\bf Landau-Michel principle.}\par\noindent {\it Landau theory can be worked out in the $G$-orbit space $\Om := M / G$.}

\subsection{Thermodynamic stability}

At this point we can already briefly discuss how the request of thermodynamic stability, i.e. convexity (see the introduction) is reflected in the polynomial $\^\pi (J)$.

Let us first of all consider the coregular case; now $\^\pi : \R^r \to \R$, and the $J_a$ can be considered as independent variables. The minimal Landau polynomial $\Phi (x) = \pi (J)$ will be quadratic in the $J$, and the stability is ensured by requiring that the matrix of second derivatives $D_{ih} = \pa^2 \pi / (\pa J_i \pa J_h )$ is positive definite.

So the prescription in this case will be to {\it consider a polynomial of order $N = 2 {\rm max} (d_1,...,d_r) = 2 d_r$} (recall $d_1 \le d_2 \le ... \le d_r$); and of course {\it choose coefficients so that the matrix $D$ is positive definite} for large $|x|$.

Note that if we deal with a non-coregular case this prescription also works: maybe it would be possible to stop at a lower order, as we have to care only about the submanifold of $\Om$ allowed by the relations between the $J_a$, but if we require stability in all of $\Om$ we are on the safe side.

We stress that the prescription is {\it not } to write $\pi$ as a quadratic polynomial in the $J_a$ and then express $\Phi$ in terms of this; rather it is to {\it consider the most general $G$-invariant polynomial of order $2 d_r$}. This can contain quite high powers in some of the $J_a$'s, see example 6.

It is essential to recall that the coefficients of (at least some of) the polynomials will depend on the external parameters; in particular, this will be the case for $J_1 = |x|^2$, whose coefficient controls the loss of stability of the critical point $x = 0$ and thus the onset of the phase transition.

\medskip\noindent
{\bf Example 4} {\it (continued).} In the case of example 4, the basic invariants are both quadratic, so we consider an expansion quadratic in these. That is, we consider
$$ \begin{array}{l}
\Phi \ = \ \Phi_1 \, + \, \Phi_2 \ ; \\
\Phi_1 \ = \ a_1 J_1 + a_2 J_2 \ , \\
\Phi_2 \ = \ (b_1 J_1^2 + b_2 J_2^2) \, + \, c_1 J_1 J_2 \ . \end{array} $$
Thus $\Phi$ depends on 5 coefficients. The origin is stable provided both the $a_i$ are positive; thus (at least one of) the $a_k$ should depend on parameters, and pass through zero, in order to have a phase transition. This example will be studied in section 4 below.
\EOE

\medskip\noindent
{\bf Example 5} {\it (continued).} In the case of Example 5, all basis polynomials are quadratic in $(x,y)$, so we have to consider again an expansion up to order four in $(x,y)$, hence quadratic in the $J_\a$.
This means we will consider
$$ \begin{array}{l}
\Phi \ = \ \Phi_1 \, + \, \Phi_2 \ ; \\
\Phi_1 = (a_1 J_1 + a_2 J_2 + a_3 J_3) \ , \\
\Phi_2 = \[ (b_1 J_1^2 + b_2 J_2^2 + b_3 J_3^2) \, + \, (c_1 J_1 J_3 + c_2 J_1 J_2 ) \] \ . \end{array} $$
We have not included terms of the form $c_0 J_1 J_2$ in $\Phi_2$: as $J_1 J_2 = J_3^2$, these can be absorbed in $b_3 J_3^2$.
Hence $\Phi$ depends on 8 coefficients; note that again (at least some of) the $a_k$ should depend on parameters to have a phase transition. This example will also be studied in section 4 below.
\EOE

\medskip\noindent
{\bf Example 6} {\it (continued).} In the case of Example 6, we have $d_1 = 2$, $d_2 = 4$, $d_3 = 6$. Hence we have to consider an expansion up to order 12 in $(x,y,z)$. We write
$$ \Phi \ = \ \Phi_1 + \Phi_2 + \Phi_3 + \Phi_4 + \Phi_5 + \Phi_6 $$
and the homogeneous terms $\Phi_m$ (of order $2m$ in $x,y,z$) are given by
$$ \begin{array}{l}
\Phi_1 \ = \ a_1 J_1 \\
\Phi_2 \ = \ a_2 J_2 + b_1 J_1^2 \\
\Phi_3 \ = \ a_3 J_3 + b_2 J_1^3 + c_1 J_1 J_2 \\
\Phi_4 \ = \ b_3 J_1^4 + b_4 J_2^2 + c_2 J_1^2 J_2 + c_3 J_1 J_3 \\
\Phi_5 \ = \ b_4 J_1^5 + c_4 J_1^3 J_2 + c_5 J_1^2 J_3 + c_6 J_1 J_2^2  + c_7 J_2 J_3 \\
\Phi_6 \ = \ b_5 J_1^6 + b_6 J_2^3 + b_7 J_3^2 + c_8 J_1^4 J_2 + c_9 J_1^3 J_3 + c_{10} J_1^2 J_2^2 + c_{11} J_1 J_2 J_3 \ . \end{array} $$
Thus $\Phi$ depends on 21 coefficients; it is obvious that we need to simplify it in order to be able to discuss its behaviour.

In \cite{SGU}, Sergienko, Gufan and Urazhdin argued that in this case one can just consider
$$ \Phi = \a_1 J_1 + \b_1 J_2 + \ga_1 J_3 + \a_2 J_1^2 + \b_2 J_2^2 + \ga_2 J_3^2 \ , $$
i.e. a potential quadratic in the basic invariants and with no mixed terms (terms of the type $J_i J_k$ with $i \not= k$); this was further justified and considered in \cite{SGU}.

Note that the stability of the origin is controlled by the sign of the coefficient $a_1$: so this should depend on parameters, and pass through zero, in order to have a phase transition.

This example will be studied in detail in section 5 below.
\EOE

\section{Poincar\'e and Lie-Poincar\'e transformations}
\def\sn{2}

In this section we recall a technique which is the fundamental tool of  the theory of Poincar\'e-Birkhoff normal forms \cite{ArnG,Elp,Wal,CGs,Gae02}, i.e. {\it Poincar\'e  transformations} and an improved version of these, {\it Lie-Poincar\'e transformations}.

Were we dealing with the complete Taylor expansion $\^\Phi (x)$, we would need the whole normal form theory; as we only consider a truncated polynomial $\Phi (x)$, we will only need some part of the theory.

\bigskip

Consider a neighbourhood $D$ of the origin in $\R^n$ with coordinates $(x^1, ... , x^n )$; and the Taylor series for a smooth function $F(x) \in \R$, vanishing in the origin:
$$ F (x) \ = \ \sum_{k=0}^\infty \, F_k (x) \ \ ; \ \ F_k (a x) = a^{k+1} \, F(x) \ . \eqno(\sn.1) $$
Note that $F_k$ is homogeneous of degree $k+1$: the reason for this notation will be clear below. We will denote the set of smooth scalar (vector) functions homogeneous of degree $k+1$ as $S_k$ ($V_k$).

We want to consider near-identity changes of coordinates in $D$ generated by a homogeneous vector function $h_k \in V_k$ with  components $h_m^i \in S_k$ (Poincar\'e transformations), and how these are reflected in the expansion (\sn.1) for $F$.

If we write the change of coordinates in the form
$$ x^i \ = \ y^i + h^i_m (y) \ \ ; \ \ h^i_m \in S_m \eqno(\sn.2) $$
then we have (using an obvious simplified notation)
$$ F_k (x) = F_k (y + h) = F_k (y) + \sum_{j=1}^n {\pa F_k \over \pa y^j} h^j + \sum_{i,j=1}^n {1 \over 2} {\pa^2 F \over \pa y^i \pa y^j } h^i h^j + ... \eqno(\sn.3) $$
Here the term with with $q$ derivatives ($q=0,1,...$) belongs to $S_{k+qm}$, with $q = 0,1,....$.

This shows at once how the $F_k$ are transformed; if we write $F(x) = \^F (y) = \sum_k \^F_k (y)$, then under (\sn.2) we have
$$ \cases{\^F_k = F_k & for $k < m$ \cr \^F_k = F_k + (h_m \cdot \grad) F_{k-m} & for $m \le k < 2m$ \cr \^F_k = F_k + (h_m \cdot \grad) F_{k-m} + (1/2) h_m^2 \triangle F_{k-2m} & for $2m \le k < 3m$ \cr ...... & ....... \cr} \eqno(\sn.4) $$

A modification of this technique, more suited to our needs, is known as Lie-Poincar\'e (LP) transformations \cite{Dep,MPL,Wal,BGG}. With this, rather than considering the change of coordinates (\sn.2), we will consider the change of coordinates obtained as the time one flow of the vector field $ {\dot x}^i = h^i_m (x)$: if $\Phi(t;x_0)$ is the flow under this with initial condition $x(0) = x_0$ then we define the change of coordinates as $ x = \Phi (1;y)$, also written as $x = e^h y$.

Needless to say, this just coincides with (\sn.2) at first order; however, this approach has several advantages. Leaving apart the theoretical ones \cite{BGG}, we stress that LP transformations are better suited to deal with the case where the function $F$ has some symmetry property which we want to preserve in the change of coordinates.

The last statement is better understood, anticipating slightly our discussion, considering the gradient $f = \grad F$; we write $f = \sum f_k$, with $f_k = \grad F_{k+1} \in V_k$.

Under a LP change of coordinates we have (see e.g. \cite{CGs,Gae1} for details)
$$ \^f_k \ = \ \sum_{s=0}^{[k/m]} \ {1 \over s!} \ \h_m^s (f_{k - s m}) \ ; \eqno(\sn.5) $$
here we have denoted by $[a]$ the integer part of $a$, and introduced the operator $\h_m$ defined by
$$ \h_m (f) \ := \ \{ h_m , f \} \ := (h_m \cdot \grad) f - (f \cdot \grad) h_m \ . \eqno(\sn.6) $$
Note that (\sn.5) is just the Baker-Campbell-Haussdorf formula\footnote{Here the role of the commutator is taken by the bracket $\{.,.\}$, which is nothing else than a translation of $[.,.]$ in terms of components of the commutator $[X_h , X_f ]$, with $X_h = h^i \pa_i$ and $X_f = f^i \pa_i$. Indeed, for $X_h$ and $X_f$ as above,  $[X_h,X_f] = \{ h , f \}^i \pa_i$.}; it states that the vector function $f (x)$ is written as $\^f (y)$ with $$ \^f \ = \ \[ e^{th} \, f \, e^{-th} \]_{t=1} \ . \eqno(\sn.7) $$

Let us now briefly comment on how these changes of coordinates are used (we keep to the vector case for ease of discussion); this rests on (\sn.4).

With both the Poincar\'e and Lie-Poincar\'e approaches, if we consider a change of coordinates with generator $h_m \in S_m$, then: $(a)$ the terms of degree $k < m$ are not changed at all; and $(b)$ the terms of degree $m$ (actually, of degree $m \le k < 2m$) are changed in a very simple way, depending only on the linear part of the system, i.e. as
$ \^F_m = F_m + \h_m (f_0)$. (Higher degree terms change in a slightly different way depending if we are using the Poincar\'e or Lie-Poincar\'e approach).

Property $(a)$ allows to consider sequentially changes of coordinates with generating functions homogeneous of degree $2,3,....$; each time one is not changing terms of lower degree -- so that terms in $S_k$ can be considered as stable after step $k$ -- and by property $(b)$ one is able to eliminate all terms in the range of the operator $\L_0 \ := \{ f_0 , . \}$. Terms in the complementary space cannot be eliminated, and are said to be {\it resonant}.

Hence, by applying repeatedly changes of coordinates of this type, one can reach a system which is in normal form -- namely, contains only resonant terms -- up to any desired order $N$.

\section{Transformation of invariant polynomials}
\def\sn{3}

\subsection{Lie-Poincar\'e transformations and invariant polynomials}

Let us apply the technique illustrated in the previous section to $G$-invariant polynomials $\Phi (x) = \pi [J (x)]$. We write
$$ \Phi (x) \ = \ \sum_{k=0}^\infty \, \Phi_k (x) \eqno(\sn.1) $$
where $\Phi_k (a x) = a^{k+1} \Phi (x)$.

We want to consider changes of coordinates of the form
$$ x^i \ = \ y^i \ + \ h^i (y) \ ; \eqno(\sn.2) $$
moreover, we want to preserve the symmetry properties of $\Phi$: the change of coordinates should be $G$-equivariant. Thus the function $h: M \to M$ has to transform in the same way as $y$ under the $G$-action, i.e. we have to require that
$$ h (T_g y) \ = \ T_g \, h (y) \ \ \ \ \ \forall y \in M \ \forall g \in G \ . \eqno(\sn.3) $$

We will choose $h$ to be the gradient of a $G$-invariant function $H (x)$, i.e.\footnote{For general $M$ with metric $g$, the $\delta$ in the formula (\sn.4) should be replaced by $g$.}
$$ h^i (y) \ = \ \delta^{ij} \ { \pa H (y) \over \pa y^j} \ . \eqno(\sn.4) $$

We stress that in general there can be equivariant vector polynomials which are not obtained as the gradient of invariant scalar polynomials\footnote{This can be the case for very simple groups: it happens e.g. for any representation of $SO(3)$ at the exception of the fundamental one, and $h$ of degree higher than 3; see \cite{Gold} for details.}; however, the choice (\sn.4) guarantees that (\sn.3) is satisfied: we are not as general as possible, but we are on the safe side.

In order to know how (\sn.2) acts on (\sn.1), it suffices to know how it acts on the basic invariants $J_a$. This action is simply
$$ \begin{array}{rl}
J_a (x) \ =& \ J_a [y + h(y)] \ = \ J_a (y) + (\de J_a) (y) \\
& = \ J_a (y) + (\pa J_a / \pa y^p) h^p + {1 \over 2} (\pa^2 J_a /  \pa y^p \pa y^q) h^p h^q + ... \end{array} \eqno(\sn.5) $$
and the action on $\Phi$ is therefore
$$ \begin{array}{rl}
\Phi (x) =& \pi [J_1 (x),...,J_r (x)] = \cr
=& \pi [J_1 (y) , ... , J_r (y) ] + \sum_{a=1}^r (\pa \pi / \pa J_a) \, [(\de J_a)(y)] \\
& + \, (1 / 2) \, \sum_{a,b=1}^r \, (\pa^2 \pi / \pa J_a \pa J_b) \, [(\de J_a)(y)] \, [(\de J_b) (y)] + .... \ \ . \end{array} \eqno(\sn.6) $$

Note that if $\Phi$ is of finite order $N$, this change of coordinates will in general produce terms of higher order, i.e. $\Phi$ will be changed into a polynomial of order higher than $N$. Thus, the truncation to order $N = 2 d_r$ should be performed again after all the required Poincar\'e changes of coordinates took place (or after each one, for computational convenience).

The formulas (\sn.5), (\sn.6) can be considerably involved, and in principles can be read off equation (2.5). Luckily, for our discussion we only need the first order terms; dropping higher order terms, (\sn.5) reads
$$ J_a (x) = J_a (y) + (\pa J_a / \pa y^p) h^p \ ; \eqno(\sn.7) $$
recalling (\sn.4), this is
$$ J_a (x) = J_a (y) + (\pa J_a / \pa y^p) \de^{pq} (\pa H / \pa y^q) \ . \eqno(\sn.8) $$

As $H$ is $G$-invariant, it is also possible to write it as a function of the basic invariants: $H(x) = \chi [J_1 (x) , ... , J_r (x)]$. Hence,
$$ {\pa H \over \pa y^k} \ = \ {\pa H \over \pa J_b} \ \cdot \ {\pa J_b \over \pa y^k} \ , \eqno(\sn.9)$$
and (\sn.8) reads
$$ \begin{array}{rl}
J_a (x) \ =& \  J_a (y) \, + \, (\pa J_a / \pa y^p) \, \de^{pq} \, (\pa H / \pa J_b) \, (\pa J_b / \pa y^q) \\
& = \ J_a (y) \ + \
\sum_{b=1}^r \, \< \grad J_a | \grad J_b \> \, (\pa H / \pa J_b) \ . \end{array} \eqno(\sn.10) $$

If now we recall (1.2), the above can be rewritten as
$$ \begin{array}{l}
J_a (x) \ = \ J_a (y) \ + \ (\de J_a) (y) \ ;
\de J_a \ := \ \P_{\a \b} \, (\pa H / \pa J_b ) \ . \end{array}  \eqno(\sn.11) $$

\subsection{Elimination of terms in invariant polynomials}

Let us now apply the above discussion to the reduction of an invariant polynomial $\Phi (x) = \Psi (J_1 , ... , J_r )$. We have in general
$$ \Psi (J) \ \to \ \Psi (J + \de J) \ = \ \Psi (J) + \sum_{\a=1}^r \, {\pa \Psi (J) \over \pa J_\a} \, \de J_\a \ + \ {\rm h.o.t.} \ . \eqno(\sn.12) $$
Under a change of coordinates of the form (\sn.2), with $h$ given by (\sn.4), the $J_a (x)$ change according to (\sn.11). We will write $D_\a := \pa / \pa J_\a$, and understand summation over repeated indices is implied; disregarding higher order terms and using the explicit form of (\sn.11), we get from (\sn.12) that
$$ \de \Psi \ = \ {\pa \Psi \over \pa J_\a} \, \P_{\a \b} \, {\pa H \over \pa J_\b} \ \equiv \ (D_\a \Psi ) \, \P_{\a \b} \, (D_\b H ) \ . \eqno(\sn.13) $$

Let us now consider the expansion (\sn.1) for $\Phi$, and hence for $\Psi = \sum_k \Psi_k$, where $\Phi_k (x) := \Psi_k (J)$.

Here we are introducing a grading with respect to the $x$ coordinates; note that $J_a$ is of degree $d_a$, hence $\grad J_a$ is of degree $(d_a-1)$, and the element $\P_{\a \b}$ of the $\P$-matrix is of degree $(d_\a + d_\b - 2)$. We denote by $d_h = m+1$ the degree of the function $h$ (which is $d_H -1$, with $d_H = m+2$ the degree of the scalar function $H_m$).
Thus the element $(D_\a \Psi_k ) \, \P_{\a \b} \, (D_\b H_m )$ (no sum on $\a , \b)$ is of degree
$$ (k+1 - d_\a) + (d_\a + d_\b - 2) + (d_H - d_\b) \ = \ k + d_H -1 \ = \ k + m +1 \ , $$
i.e. $(D_\a \Psi_k ) \P_{\a \b} (D_\b H_m ) \in S_{m+k} $.

This means that under a change of coordinates (\sn.2) generated by $H_m$ the terms $\Psi_k$ with $k \le m$ are not changed, while the terms $\Psi_{m+p}$ change according to
$$ \Psi_{m+p} \ \to \^\Psi_{m+p} = \Psi_{m+p} + (D_\a \Psi_p ) \P_{\a \b} (D_\b H_m ) \ + \ {\rm h.o.t.} \ . \eqno(\sn.14) $$

We can then proceed as sketched in section 2, i.e. operate sequentially with $H$ in $S_1$, $S_2$, .... ; at each stage (generator $H_m$) we are not affecting the terms $\Psi_k$ with $k \le m$. Moreover, we can just consider the first order correction, as higher order terms are generic and will be taken care of in subsequent steps.

Consider now, to fix ideas, the case where $d_1 = d_2 = ... = d_s = 2$ (recall this implies there are no invariants of order less than two).
Then $\Psi_{m+1}$ changes according to
$$ \Psi_{m+1} \ \to \^\Psi_{m+1} = \Psi_{m+1} + (D_\a \Psi_1 ) \P_{\a \b} (D_\b H_m ) \ + \ {\rm h.o.t.} \ . \eqno(\sn.15) $$
If $\Psi_1 \not= 0$, we disregard higher order terms; note that necessarily $\Psi_1 = a_k J_k$ (the sum on $k$ runs from 1 to $s$). In this way we obtain that all terms which are of the form
$$ \sum_{k=1}^s \sum_{\b=1}^r a_k P_{k \b} (D_\b H_m ) \eqno(\sn.16) $$
for $H \in S_{m+1}$ can be eliminated by our procedure. Needless to say, if the $a_k$ are all zero we need to consider higher order terms.

Note also that in order to perform this procedure, we should determine suitable generating functions $H_m$; this can be done by requiring as many component of $\^\Psi_{m+p}$ as possible to vanish, and inverting $(D_\a \Psi_p) \P_{\a \b}$ to solve (3.14) for $H_m$. The relevant point is, that this requires $D_\a \Psi_p \not= 0$, and this condition could fail for certain values of the parameters defining $\Psi_p$ in terms of the basis invariant polynomials.

Another relevant point should also be addressed: eqs. (3.13) and (3.16) say that, in this context (see the discussion below), the functions $Q_i$ appearing in (0.2) should be identified with the derivatives $D_i H_*$ for a suitable function $H_*$ (this will be obtained from the generating functions $H_m$ used at each step). However, this identificiation leads to a natural ``compatibility condition'' between the $Q_i$'s: that is, we have to require 
$$ D_i \, Q_j \ = \ D_j \, Q_i \ . \eqno(\sn.17) $$

\medskip\noindent
{\bf Example 4} {\it (continued).} In the case of example 4, the two basic invariants $J_1 = x^2$ and $J_2 = y^2$ are both quadratic, and thus contribute to the first order formulas (\sn.13), (\sn.15).

Let us consider a quartic generating function $H_2 = h_1 J_1^2 + h_2 J_2^2 + k_1 J_1 J_2$; the quartic term $\Psi_2 = b_1 J_1^2 + b_2 J_2^2 + c J_1 J_2$ will change to $\Psi_2 + \de \Psi_2$, and from our formulas we get
$$ \de \Psi_2 \ = \ (8 a_1 h_1 ) \, J_1^2 \, + \, (8 a_2 h_2 ) \, J_2 \, + \, [4 (a_1 + a_2) k_1 ] \, J_1 J_2 \ . $$
Therefore, provided the conditions
$$ a_1 \not= 0 \ , \ a_2 \not= 0 \ , \ a_1 + a_2 \not= 0 $$
are verified, we can choose $h_1 , h_2 , k_1$ such that $\^\Psi_2 = \Psi_2 + \de \Psi_2 = 0$. These are given explicitly by
$$ h_1 = - {b_1 \over 8 a_1 } \ , \ h_2 = - {b_2 \over 8 a_2 } \ , \ k_1 = - {c \over 4 (a_1 + a_2 ) } \ . $$

Note that actually we do not want to set the term $\^\Psi_2$ to zero, in order to satisfy the requirement of thermodynamic stability. Thus this is just an example of what could be done by the Poincar\'e procedure, but does not apply directly to Landau theory.
\EOE

\medskip\noindent
{\bf Example 5} {\it (continued).} In the case of Example 5, again all the basic invariants are quadratic in $(x,y)$. Thus all of them contribute to the first order formulas (\sn.13), (\sn.15). Recall in this case the invariants satisfy the relation $J_3^2 = J_1 J_2$.

We consider a quartic generating function, which we write as
$$ H_2 \ = \ h_1 J_1^2 + h_2 J_2^2 + h_3 J_3^2 + k_1 J_1 J_3 + k_2 J_2 J_3 \ . $$
Under the change of coordinates generated by this, the quartic term in the Landau polynomial, $ \Psi_2 = b_1 J_1^2 + b_2 J_2^2 + b_3 J_3^2 + c_1 J_1 J_3 + c_2 J_2 J_3$, will change to $\^\Psi_2 = \Psi_2 + \de \Psi_2$.

It follows from our general formulas, with standard algebra and using the relation $J_3^2 = J_1 J_2$, that
$$ \begin{array}{rl}
\de \Psi_2 =& (8 a_1 h_1 + a_3 k_1) J_1^2 + (8 a_2 h_2 + a_3 k_2) J_2^2 + [4 (a_1 + a_2) h_3 + 3 a_3 (k_1 + k_2 ) ] J_3^2 + \\
&  + [4 a_3 h_1 + 2 a_3 h_3 + (6 a_1 + 2 a_2 ) k_1 ] J_1 J_3 +
[4 a_3 h_2 + 2 a_3 h_3 + (2 a_1 + 6 a_2 ) k_2 ] J_2 J_3 \ . \end{array} $$

Thus, we can choose $\{h_i , k_i \}$ such that $\^\Psi_2 = 0$\footnote{Again, one does not really want this in the context of Landau theory, as it would violate the requirement of thermodynamic stability; see the previous example.} if the matrix
$$ \pmatrix{
8 a_1 & 0 & 0 & a_3 & 0 \cr
0 & 8 a_2 & 0 & 0 & a_3 \cr
0 & 0 & 4 (a_1 + a_2) & 3 a_3 & 3 a_3 \cr
4 a_3 & 0 & 2 a_3 & (6 a_1 + 2 a_2) & 0 \cr
0 & 4 a_3 & 2 a_3 & 0 & (2 a_1 + 6 a_2) \cr} $$
has nonvanishing determinant. In fact, with obvious notation, $\^\Psi_2 = M ({\bf h},{\bf k}) + ({\bf b},{\bf c})$. It results
$$ \begin{array}{rl}
|M| \ =& \ 256 \, [ 12 a_1^4 a_2 - 3 a_2^3 a_3^2 + a_2 a_3^4 + a_1^3 (52 a_2^2 - 3 a_3^2) \\
 & \ \ + a_1^2 (52 a_2^3 - 17 a_2 a_3^2) + a_1 (12 a_2^4 - 17 a_2^2 a_3^2 + a_3^4) ] \ . \end{array} $$

Let us discuss the vanishing or otherwise of this. If any two of the $a_i$ are zero, then $|M|=0$. If $a_1 = 0$ (and of course $a_2 \not= 0 \not= a_3$), then $|M| = (a_2 a_3^2) (a_3^2 - 3 a_2^2)$, i.e. we can still set $\^\Psi_2 = 0$ provided $a_3 \not= \pm \sqrt{3} a_2$. Similarly, if $a_2 = 0$ (and of course $a_1 \not= 0 \not= a_3$), then $|M| = (a_1 a_3^2) (a_3^2 - 3 a_1^2)$, and we can set $\^\Psi_2 = 0$ provided $a_3 \not= \pm \sqrt{3} a_1$. For $a_3 = 0$ (and of course $a_1 \not= 0 \not= a_2$), we get
$$ |M| \ = \ 4 a_1 a_2 (3 a_1^3 + 13 a_1^2 a_2 + 13 a_1 a_2^2 + 3 a_2^3) \ ; $$
this is nonzero provided $a_1 \not= - a_2$, $a_1 \not=  -3 a_2$, $a_1 \not= - a_2/3$.

Finally, if the $a_i$ are all nonzero, the determinant is nonzero provided all of the following conditions are satisfied:
$$ \cases{
a_1 \not= - a_2 & \cr
a_3^2 \not= 4 a_1 a_2 & \cr
a_3^2 \not= 3 a_1^2 + 10 a_1 a_2 + 3 a_2^2 & \cr} $$
We won't pursue the analysis of the degenerate cases; the point here is to show that all the computations can be done in complete detail using only standard algebra.
\EOE

\medskip\noindent
{\bf Example 6} {\it (continued).} In the case of Example 6, only $J_1$ is quadratic, so first order formulas involve only this, and cannot be used if $a_1 = 0$.

We will postpone discussion of this example until sect.5, where we consider it in the more general case of varying control parameters, i.e. varying coefficients $a_i$, $b_i$, $c_i$.
\EOE
\medskip

We stress that in examples 4 and 5, we could use the simplified formulas (3.13), (3.15), based on linear action (quadratic generating function). However, in general we should consider the full change of coordinates generated by $H$. Note that in this case the transformation eliminating terms of a given order will at the same time generate terms of higher orders.

\subsection{The simplifying criterion, and discussion.}

The discussion conducted in this and the previous section provides  rigorous ground for the symplifying criterion stated in the Introduction and due to Gufan.

It should be stressed that here we have worked with a given Landau polynomial, i.e. with {\it fixed} values of the parameters entering in it. We stress that these parameter -- or at least some of them -- will in general depend on the external ``control'' parameter (temperature, pression, magnetic field, etc), and indeed they have to change with these for a phase transition to take place.

Thus our discussion so far provides a proof of the Gufan simplifying criterion only when we work at a given value of the control parameter(s); we will see in the next section that some extra care is needed if we want to work on a full interval of values of the control parameter(s).

This is reflected in our statement of the simplifying criterion, see the Introduction, as there we consider a given potential with no mention of dependence upon external parameters.

Another relevant remark should be made. In this section, and in the whole paper, we consider just changes of coordinates: that is, we eliminate terms by choosing suitable coordinates, but we are {\it not } changing the potential; on the other hand, in singularity theory one allows changes of the potential, provided these do not alter its qualitative behaviour \cite{ArnS}. Thus, we are {\it not }  considering the most general transformation of $\Phi$ allowed by Landau theory. On the other hand, the transformations considered here are surely allowed, and actually can be easily implemented (algorithmically) via a symbolic manipulation language.

In view of the above, it is worth stating a weaker form of the simplifying criterion given in the introduction. This is the result being actually proven by our discussion so far; note that it should be called a ``reduction'' -- rather than ``simplifying'' -- criterion: indeed, as already remarked, we do not simplify the potential, but just provide a reduced expression of it by using more convenient (local) coordinates.

\medskip\noindent
{\bf Reduction criterion.}
{\it Let $G$ be a compact Lie group, acting in $\R^n$ through a linear representation; let $\{ J_1 , ... , J_r \}$ be a minimal integrity basis for $G$, and let $F(J_1,...,J_r) : \R^n \to \R$ be a potential. Define, for $i=1,...,r$ and with $(.,.)$ the scalar product in $\R^n$, the quantities
$$ U_i (J_1,...,J_r) \ := \ \sum_{k=1}^r \, {\pa F \over \pa J_k} \ (\grad J_k , \grad J_i ) \ . \eqno(\sn.18) $$
Let $B \sse \R^n$ be a sufficiently small neighbourhood of the origin in $\R^n$. Then there is a sequence of Poincar\'e changes of coordinates in $B$, such that the potential $F$ is expressed in an equivalent form $\^F$, where terms of $F$ which can be written as
$$ \sum_{\a=1}^r \ Q_\a (J_1 , ... , J_r) \ U_\a (J_1 , ... , J_r ) \ + \ {\rm h.o.t.} \ , \eqno(\sn.19) $$
with $Q_\a$ polynomials in $J_1,...,J_r$ satisfying the compatibility condition $$ {\pa Q_\b \over \pa J_\a} \ = \ {\pa Q_\a \over \pa J_\b} \eqno (\sn.20) $$ and ``h.o.t.'' denoting higher order terms, can be eliminated in $\^F$.}
\medskip

The ``sufficiently small'' in the statement above should be meant in the sense that the overall Poincar\'e change of coordinates described by the combination of the different ones at each order $k = 2,..., 2 d_r$ should define a convergent series. In general -- for a given finite order $d_r$ -- this will be the case only in some neighbourhood $B$ of the origin. In this respect, it should be stressed that the sutuation considered here is naturally characterized by a symmetry, hence the symmetric theory (due to Markhashov, Bruno, Walcher and Cicogna) should be used\footnote{We will not discuss this theory here, and just refer e.g. to the discussion given in \cite{BrW,CiW}.}. Note however that when one performs a concrete computation, the radius of convergence will always be immediately read off the concrete (denominators in the) inversion formulas.

\section{Reduction criterion with varying parameters}
\def\sn{4}

In Landau theory, one is considering (the vicinity of) phase transitions; that is, the coefficients of the polynomial $\Phi (x)$ depend on external control parameters $\la$, and necessarily pass through critical values.

In this section, we investigate how the discussion given so far should be modified if we want to consider not just given fixed values of the parameter(s), i.e. of the coefficients appearing in the Landau polynomial, but a full range of values, including in particular critical ones.

Let us, to fix ideas, consider $\la \in \R$ and let $J_1 = |x|^2$ be the only quadratic basic invariant, so that $\Phi = c_1 (\la ) J_1 + c_2 (\la ) J_1^2$ (with $c_2 (\la ) > 0$) and the loss of stability of the $x=0$ critical point is controlled by the sign of the coefficient $c_1 (\la)$ of $J_1$ in $\Phi$ -- say $c_1 (\la) = - \la /2$ with standard normalization -- so that the transition occurs at $\la = 0$,  with the symmetry-breaking phase appearing for $\la > 0$ (this is e.g. the case for $\Phi = - \la |x|^2/2 + |x|^4/4$).

As we want to describe a small but finite interval of values of $\la \in \La := [\eps_- , \eps_+]$, with $\eps_- \le 0 $ and $\eps_+ > 0$, we have to require that the near-identity changes of variables considered in previous sections are defined uniformly in $\La$. This means that they must be well defined also at $\la = 0$; in particular, changes of variables requiring a division by $c_1 (\la )$ are {\it not} allowed.

More generally (i.e. for general degrees of polynomials), we are allowed to consider only those Poincar\'e and Lie-Poincar\'e transformations which are {\it smooth and well defined in a full neighbourhood of the critical point, and in particular at the critical point itself}.

This means that the reduction criterion stated in the previous section  should be suitably restricted, as follows.

\medskip\noindent
{\bf Reduction criterion with varying parameters.}
{\it Let $G$ be a compact Lie group, acting in $\R^n$ through a linear representation; let $\{ J_1 , ... , J_r \}$ be a minimal integrity basis for $G$, and let $F(\la ; J_1,...,J_r) : \La \times \R^n \to \R$ be a potential depending on the control parameters $\la \in \La \sse \R^p$. Define, for $i=1,...,r$ and with $(.,.)$ the scalar product in $\R^n$, the quantities
$$ U_i (\la;J_1,...,J_r) \ := \ \sum_{k=1}^r \, {\pa F \over \pa J_k} \ (\grad J_k , \grad J_i ) \ . \eqno(\sn.1) $$
Let $B \sse \R^n$ be a sufficiently small neighbourhood of the origin in $\R^n$. Then there is a sequence of Poincar\'e changes of coordinates in $B$, such that the potential $F$ is expressed in an equivalent form $\^F$ in a neighbourhood $\Lambda$ of the critical value $\la = \la_0$ in the space of the control parameters $\la$; terms of $F$ which can be written {\rm uniformly in $\La$}, as
$$ \sum_{i=1}^r \ Q_i (J_1 , ... , J_r) \ U_i (\la;J_1 , ... , J_r ) \ + \ {\rm h.o.t.} \eqno(\sn.2) $$
where $Q_\a$ are polynomials in $J_1,...,J_r$ satisfying the compatibility condition $ (\pa Q_\a / \pa J_\b) = (\pa Q_\b / \pa J_\a)$ and ``h.o.t.'' denotes higher order terms, can be eliminated in $\^F$.}
\medskip

Note the only difference with respect to the criterion given in the previous section lies in the dependence on $\la$ and in requiring uniformity in $\La$, as emphasized in the statement.

In order to better understand this point, it is useful to consider again in detail the elementary example 4; we will then also analyze in detail example 5, while example 6 is analyzed in the next section.

\medskip\noindent
{\bf Example 4} {\it (continued).} In the case of example 4, a  quadratic generating function $H_1$ would produce a linear change of coordinates (we are not interested in these), so we start by considering a quartic generating function $H_2 = \b_1 J_1^2 + \b_2 J_2^2 + \b_3 J_1 J_2$. This does not affect $\Phi_1$, while $\Phi_2$ is changed into $\^\Phi_2 = \Phi_2 + \de \Phi_2$ where, according to our general formula (3.15),
$$ \de \Phi_2 \ = \ (D \Phi_1) \P (D H_2) \ = \ 8 (a_1 \b_1 J_1^2 + a_2 \b_2 J_2^2 ) + 4 (a_1 + a_2) \b_3 J_1 J_2 \ . $$
With the obvious notation $\^\Phi_2 = \^b_1 J_1^2 + \^b_2 J_2^2 + \^b_3 J_1 J_2$ and with $M$ the diagonal matrix
$ M = {\rm diag}(2 a_1 , 2 a_2 , a_1 + a_2)$, we can write in vector notation
$$ \^b \ = \ b \, + \, M \b \ .  $$
It follows immediately from this that if $a_1 a_2 \not= 0$ and $a_1 \not= - a_2$, we can set $\^b = 0$ by choosing $ \b = - M^{-1} b$; if $a_1 = 0$ but $a_2 \not= 0$, the terms $\^b_2$ and $\^b_3$ can be set to zero, while $\^b_1 = b_1$, while for $a_1 \not= 0$ and $a_2 = 0$ we can set to zero $\^b_1$ and $\^b_3$ while $\^b_2 = b_2$; if $a_1 = - a_2 \not= 0$, we can set to zero $\^b_1$ and $\^b_2$, and $\^b_3 = b_3$. Finally, for $a_1 = a_2 = 0$ the change of coordinates has no effect on $\Phi_2$.

Let us now consider this setting in the frame of Landau theory, say with a control parameter $\la \in \R$ and with a phase transition taking place at $\la = 0$ (for definiteness, with the origin a stable fixed point for $\la < 0$). The coefficients $(a_1,a_2;b_1,b_2,b_3)$ are now not constants, but in principles all depend on $\la$; as customary in physical considerations, we can assume the coefficients of higher order terms (i.e. the $b_i$) do not actually depend on $\la$. On the other hand, a dependence of the $a_i$ on $\la$ is physically essential.

For a phase transition to take place, it is needed that at $\la = 0$ at least one of $a_1, a_2$ vanish; indeed the matrix of second derivatives of $\Phi$ at $(0,0)$ is simply $V = {\rm diag} (2a_1,2a_2)$. Thus the above short discussion has an obvious physical meaning: we can drop a fourth order term like $b_i J_i^2$ only if the corresponding quadratic term $a_i J_i$ is positive definite throughout the range of control parameters we are considering; if the phase transition corresponds to $a_i$ passing through zero, we can not eliminate the corresponding $b_i J_i^2$ term. \EOE
\medskip

We will now consider in detail example 5 (and in the next section, example 6). The computations are elementary but already involve some lenghty intermediate formulas, which are therefore not shown.

We would like to stress that here we are only interested in discussing which terms can be set to zero in the Landau potential; the discussion of section 3 also shows how one could obtain -- if desired -- the final coefficients as explicit functions of the initial ones.

\medskip\noindent
{\bf Example 5} {\it (continued).} Recall that for Example 5 we had
$ \Phi = \Phi_1 + \Phi_2$ with $\Phi_k$ homogeneous of degree $2k$, given explicitely by $\Phi_1 = a_1 J_1 + a_2 J_2 + a_3 J_3$ and $$ \Phi_2 \ = \ b_1 J_1^2 + b_2 J_2^2 + b_3 J_3^2 + c_1 J_1 J_2 + c_2 J_1 J_3 + c_3 J_2 J_3 \ . $$
The $J$-gradients $D \Phi_k$ are easily computed;
these are $(D \Phi_1) = (a_1 , a_2 , a_3 )^T$, and $(D \Psi_2) = (2 b_1 J_1 + c_1 J_2 + c_2 J_3 , c_1 J_1 + 2 b_2 J_2 + c_3 J_3 , c_2 J_1 + c_3 J_2 + 2 b_3 J_3)^T$. The $\P$-matrix is also immediately obtained from these:
$$ \P \ = \ \pmatrix{4 J_1 & 0 & 2 J_3 \cr 0 & 4 J_2 & 2 J_3 \cr 2 J_3 & 2 J_3 & J_1 + J_2 \cr} \ . $$

Let us now consider an $H$ quadratic in the $J$; we write this as
$$ H \ = \ \b_1 J_1^2 + \b_2 J_2^2 + \b_3 J_3^2 + \ga_1 J_1 J_2 + \ga_2 J_1 J_3 + \ga_3 J_2 J_3 \ . $$
Thus the vector $(D H)$ is given by
$$ D H \ = \ \pmatrix{
2 \b_1 J_1 + \ga_1 J_2 + \ga_2 J_3 \cr
2 \b_2 J_2 + \ga_1 J_1 + \ga_3 J_3 \cr
2 \b_3 J_3 + \ga_2 J_1 + \ga_3 J_2 \cr} \ . $$

Recall that by (3.14), $\Phi_1$ will be unaffected, so we only have to compute $ \de \Phi_2 = (D \Phi_1) \P (D H_2)$.
With straightforward algebra we get
$$ \begin{array}{rl}
\de \Phi_2 \ =& \
(8 a_1 \b_1 + a_3 \ga_2) J_1^2 +
(8 a_2 \b_2 + a_3 \ga_3) J_2^2 + \\
 & + (4 (a_1 + a_2) \b_3 + 2 a_3 (\ga_2 + \ga_3 )] J_3^2 + \\
 & +
[4 (a_1 + a_2) \ga_1 + a_3 (\ga_2 + \ga_3 ) ] J_1 J_2 + \\
 & + [(6 a_1 + 2 a_2 ) \ga_2 + 2 a_3 (2 \b_1 + \b_3 + \ga_1) ] J_1 J_3 + \\
 & +
[(2 a_1 + 6 a_2 ) \ga_3 + 2 a_3 (2 \b_2 + \b_3 + \ga_1) ] J_2 J_3 \ .  \end{array} $$

Hence, in the new coordinates (after the change of coordinates with generator $H_2$) we get
$$ \^\Phi_2 = \Phi_2 + \de\Phi_2 = \^f_1 J_1^2 + \^f_2 J_2^2 + \^f_3 J_3^2 + \^f_4 J_1 J_2 + \^f_5 J_1 J_3 + \^f_6 J_2 J_3 \ . $$
We write
$$ f_i = \cases{b_i & for $i=1,2,3$ \cr c_{i-3} & for $i=4,5,6$\cr} \ \ , \ \ \xi_i = \cases{\b_i & for $i=1,2,3$ \cr \ga_{i-3} & for $i=4,5,6$\cr}  \ ; $$
with these, and defining the matrix
$$ M \ = \ \pmatrix{
8 a_1 & 0 & 0 & 0 & a_3 & 0 \cr
0 & 8 a_2 & 0 & 0 & 0 & a_3 \cr
0 & 0 & 4 (a_1 + a_2) & 0 & 2 a_3 & 2 a_3 \cr
0 & 0 & 0 & 4 (a_1 + a_2 ) & a_3 & a_3 \cr
4 a_3 & 0 & 2 a_3 & 2 a_3 & 6 a_1 + 2 a_2 & 0 \cr
0 & 4 a_3 & 2 a_3 & 2 a_3 & 0 & 2 a_1 + 6 a_2 \cr} $$
the transformation is written in vector form as
$$ \^f \ = \ f + M \xi \ . $$

We should then investigate if some component of the vector $\^f$ can be set to zero by a suitable choice of the vector $\xi$.

This is a standard linear algebra problem, but the relevant point here is that this depends on the range of values assumed by the $a_i$; it depends on the vanishing or otherwise of the $a_i$ in the control parameter range of interest for the transition under study.

In particular, the determinant of $M$ is
$$ |M| \ = \ 2^{10} \ (a_1 + a_2)^2 \, (4 a_1 a_2 - a_3^2) \, (3 a_1^2 + 10 a_1 a_2 + 3 a_2^2 - a_3^2) \ , $$
and if this is nonzero we can set $\^f$ to zero by choosing $\xi = - M^{-1} f$. Note that here the basic invariants are not independent, which explains why it may be possible to eliminate all the quartic terms even when one of the $a_i$'s vanishes -- provided the other two satisfy some nondegeneracy condition which can be read off the above formula for $|M|$.

We recall again that quartic terms can be eliminated without affecting the physical validity of the Landau potential only if the latter remains thermodynamically stable. \EOE

\section{The Sergienko-Gufan-Urazhdin model}

In this final section we apply our discussion to the model which originated this work, i.e. the Sergienko-Gufan-Urazhdin  model for highly piezoelectric perovskites \cite{SGU}; this is an example of a model too complex to be easily dealt with on the basis of ``semi-intuitive'' considerations for the simplification of the Landau potential. We just consider reduction of the Landau potential according to our general discussion, without any consideration on the analysis of the resulting potential (for this the interested reader is referred to \cite{SGU}).

From the point of view of the present paper, the SGU model falls in the group-theoretical case considered in Example 6. We recall that in that case (with a slight change of notation for what concerns the coefficients appearing in $\Phi$) the basic invariants are
$$ J_1 = x^2 + y^2 + z^2 \ , \
J_2 = x^2 y^2 + y^2 z^2 + z^2 x^2 \ , \
J_3 = x^2 y^2 z^2 \ ; $$
hence the $\P$-matrix is
$$ \P \ = \ 4 \ \pmatrix{J_1 & 2 J_2 & 3 J_3 \cr
2 J_2 & J_1 J_2 + 3 J_3 & 2 J_1 J_3 \cr
3 J_3 & 2 J_1 J_3 & J_2 J_3 \cr} \ . $$

We write the most general Landau polynomial in the form
$$ \Phi = \sum_{k=1}^6 \, \Psi_k $$
where the $\Psi_k$ are homogeneous of degree $2k$; these are, explicitely,
$$ \begin{array}{l}
\Psi_1 = a J_1 \\
\Psi_2 = b_1 J_2 + b_2 J_1^2 \\
\Psi_3 = c_1 J_3 + c_2 J_1^3 + c_3 J_1 J_2 \\
\Psi_4 = d_1 J_1^4 + d_2 J_2^2 + d_3 J_1^2 J_2 + d_4 J_1 J_3 \\
\Psi_5 = f_1 J_1^5 + f_2 J_1^3 J_2 + f_3 J_1^2 J_3 + f_4 J_1 J_2^2 + f_5 J_2 J_3 \\
\Psi_6 = g_1 J_1^6 + g_2 J_2^3 + g_3 J_3^2 + g_4 J_1^4 J_2 + g_5 J_1^3 J_3 + g_6 J_1^2 J_2^2 + g_7 J_1 J_2 J_3 \ . \end{array} $$
Note that at the phase transition where the full $G$ symmetry is broken, necessarily $a (\la_0)=0$: thus we cannot accept transformations with coefficient where $a$ enters with a negative exponent.

In \cite{SGU}, Sergienko, Gufan and Urazhdin argue that one can consider the reduced Landau polynomial
$$ \^\Phi \ = \ \ga_1 J_1 + \ga_2 J_2 + \ga_3 J_3 + \Ga_1 J_1^2 + \Ga_2 J_2^2 + \Ga_3 J_3^2 \ ; $$
in our present notation this means setting all coefficients to zero at the exception of $\{ a , b_1 , b_2 , c_1 , d_2 , g_3 \}$.

This statement is based on the observation that (with the notation used in the present note) all the other terms lie in the range of the $\P$ matrix, as explained in appendix B of their paper \cite{SGU}.

Note that they do not discuss the case where some of the lowest order coefficients vanish: we have seen that this corresponds to working at a fixed value of the control parameter, arbitrarily near but however bounded away from the phase transition point.

\subsection{Reduction of the Landau polynomial with the Lie-Poincar\'e algorithm}

We assume that both coefficients $b_1$ and $b_2$ are nonzero.

We start with a generating function $H_2 \in S_4$, written in general as
$$ H_2 = \b_1 J_2 + \b_2 J_1^2 \ . $$
Using our general formula (3.15), we get
$$ \de \Psi_2 \ = \ (D \Psi_1) \, \P \, (D H_2) \ = \ 8 \, (a \b_2 J_1^2 + a \b_1 J_2) \ ; $$
thus around the phase transition ($a=0$) this vanishes and cannot be used to simplify the Landau polynomial. In other words, we get
$$ \^\Psi_2 = \Psi_2 + \de \Psi_2  \ = \
(b_1 + 8 a \b_1 ) J_2 \, + \, (b_2 + 8 a \b_2) J_1^2 \ := \ \^b_1 J_2 + \^b_2 J_1^2 ; $$
to set $\^b_i = 0$ we should choose $\b_i = - b_i /(8 a)$, which is singular at $\la = \la_0$ since $a (\la_0) = 0$.

\subsubsection{Terms of order six}

Consider the effect of this same change of coordinates on the term $\Psi_3$: we have
$$ \begin{array}{ll}
\^\Psi_3 \ :=& \ \Psi_3 + \de \Psi_3 \ = \ (c_1 + 12 b_1 \b_1) J_3 + \\  & + (c_2 + 4 b_1 \b_1 + 16 b_2 \b_1 + 16 b_1 \b_2) J_1 J_2 +
(c_3 + 16 b_2 \b_2) J_1^3 \ := \\
 & := \ \^c_1 J_3 + \^c_2 J_1^3 + \^c_3 J_1 J_2 \ . \end{array} $$

Thus we have
$$ \^c \ = \ c \ + \ M_c \, \b $$
with the matrix $M_c$ defined as
$$ M_c \ := \ 4 \ \pmatrix{
3 b_1 & 0 \cr
(b_1 + 4 b_2) & 4 b_1 \cr
0 & 4 b_2 \cr} \ . $$
For generic $b_k$, we can set to zero any two of the three $\^c_k$. Once we have chosen these, and therefore we have determined the values of $\b_1$ and $\b_2$, it is immediate to compute the change also in higher order coefficients, i.e. determine also the $(\^d_i , \^f_i , \^g_i)$. We stress that here ``generic'' means in particular that we need $b_1 \not= 0 \not= b_2$, and $b_1 \not= - 4 b_2$.

More generally, we can eliminate any two variables provided the determinant of the corresponding two-dimensional submatrix is nonzero: thus we have the following table of conditions for the different choices to be possible:
$$ \begin{array}{ll}
c_1 = c_2 = 0 & b_1 \not= 0 \\
c_1 = c_3 = 0 & b_1 b_2 \not=0 \\
c_2 = c_3 = 0 & (b_1 + 4 b_2) \, b_2 \not= 0 \ . \\
\end{array} $$

\subsubsection{Terms of order eight}

Let us now consider the effect of the change of coordinates with generator 
$$ H_3 \ = \ \ga_1 J_3 + \ga_2 J_1^3 + \ga_3 J_1 J_2 \ . $$
We will still write the coefficients of $\Psi_k$ as $a,b_i, c_i,....$; however we stress that these will be {\it different} from the original ones, as they have been changed in the first step, i.e. under the change of coordinates generated by $H_2$: thus it would be more precise to write them as $(a,b_i,\^c_i,\^d_i,...)$, and we omit the hat for ease of notation.

We have of course a $\de \Psi_3$ proportional to $a$ and thus of no use near the transition point. As for $\de \Psi_4$, we get
$$ \begin{array}{rl}
\de \Psi_4 \ =&
(24 b_2 \ga_2) \, J_1^4 \ + \ (8 b_1 \ga_3) \, J_2^2 \\
 & + \ (24 b_1 \ga_2 + 4 b_1 \ga_3 + 24 b_2 \ga_3) \, J_1^2 J_2 \\
 & + \ (8 b_1 \ga_1 + 24 b_2 \ga_1 + 12 b_1 \ga_3) \, J_1 J_3 \ ; \end{array} $$
hence $\^\Psi_4 = \^d_1 J_1^4 + \^d_2 J_2^2 + \^d_3 J_1^2 J_2 + \^d_4 J_1 J_3$ with, in vector notation,
$$ \^d \ = \ d \ + \ M_d \ \ga $$
where the matrix $M_d$ is given by
$$ M_d \ = \ 4 \ \pmatrix{
0 & 6 b_2 & 0 \cr
0 & 0 & 2 b_1 \cr
0 & 6 b_1 & b_1 + 6 b_2 \cr
(2 b_1 + 6 b_2) & 0 & 3 b_1 \cr} \ . $$
Considerations similar to those applying for $H_2$ and $\^\Psi_3$ apply also here.

Note that, due to the structure of the matrix $M_d$, for $b_1 \not= 0 \not= b_2$ and $b_1 \not= 6 b_2$, we can eliminate any two of $\^d_1 , \^d_2 , \^d_3$, and moreover (if $b_1 \not= - 3 b_2$) also $\^d_4$.
We have the following table for conditions required for the different choices:
$$ \begin{array}{ll}
d_2 = d_3 = d_4 = 0 & b_1^2 (b_1 + 3 b_2) \not= 0 \\
d_1 = d_3 = d_4 = 0 & b_2 (b_1 + 6 b_2) (b_1 + 3 b_2 ) \not= 0 \\
d_1 = d_2 = d_4 = 0 & b_1 b_2 (b_1 + 3 b_2) \not= 0 \end{array} $$

\subsubsection{Terms of order ten}

We pass then to consider change of variables with generator
$$ H_4 \ = \ \eta_1 J_1^4 + \eta_2 J_2^2 + \eta_3 J_1^2 J_2 + \eta_4 J_1 J_3 \ . $$
As usual, the lowest term on which it acts is $\Psi_4$, but the change is proportional to $a$ and thus vanishes at the critical point.
Let us then look at its effect on $\Psi_5$, which will be mapped to
$$ \^\Psi_5 = \^f_1 J_1^5 + \^f_2 J_1^3 J_2 + \^f_3 J_1^2 J_3 + \^f_4 J_1 J_2^2 + \^f_5 J_2 J_3 \ .  $$
In this case we get
$$ \begin{array}{rl}
\de \Psi_5 \ =& \
(32 b_2 \eta1) J_1^5 +
    (32 b_1 \eta_1 + 4 b_1 \eta_3 + 32 b_2 \eta_3) J_1^3 J_2 \\
 & + (12 b_1 \eta_3 + 8 b_1 \eta_4 + 32 b_2 \eta_4) J_1^2 J_3 \\
 & + (8 b_1 \eta_2 + 32 b_2 \eta_2 + 16 b_1 \eta_3) J_1 J_2^2 \\
 & + (24 b_1 \eta_2 + 8 b_1 \eta_4) J_2 J_3 \ ; \end{array} $$
hence we can write
$$ \^f \ = \ f \ + \ M_f \eta $$
where the matrix $M_f$ is given by
$$ M_f \ = \ 4 \ \pmatrix{
8 b_2 & 0 & 0 & 0 \cr
8 b_1 & 0 & (b_1 + 8 b_2) & 0 \cr
0 & 0 & 3 b_1 & (b_1 + 8 b_2) \cr
0 & (2 b_1 + 8 b_2) & 4 b_1 & 0 \cr
0 & 6 b_1 & 0 & 2 b_1 \cr} \ . $$
\def\th{\vartheta}
The usual considerations apply here: we can in general eliminate four out of the five coefficients, and the table giving conditions for each choice is the following:
$$ \begin{array}{ll}
f_2 = f_3 = f_4 = f_5 = 0 & b_1 (3 b_1 + 20 b_2) \not= 0 \\
f_1 = f_3 = f_4 = f_5 = 0 & b_1 b_2 (3 b_1 + 20 b_2) \not= 0 \\
f_1 = f_2 = f_4 = f_5 = 0 & b_1 b_2 (b_1 + 4 b_2^2 ) (b_1 + 8 b_2^2) \not= 0 \\
f_1 = f_2 = f_3 = f_5 = 0 & b_1 b_2 (b_1 + 8 b_2^2) \not= 0 \\
f_1 = f_2 = f_3 = f_4 = 0 & b_2 (b_1 + 4 b_2^2 ) (b_1 + 8 b_2^2) \not= 0 \end{array} $$

\subsubsection{Terms of order twelve}

Finally, let us consider a change of coordinates with generator
$$ H_5 \ = \ \th_1 J_1^5 + \th_2 J_1^3 J_2 + \th_3 J_1^2 J_3 + \th_4 J_1 J_2^2 + \th_5 J_2 J_3 \ ; $$
again we write the coefficients of the $\Phi_k$ with no hat for ease of notation. As we know, the only two terms affected by this will be $\Phi_5$ (with $\de \Phi_5$ proportional to $a$ and thus of no use) and
$$ \Psi6 \ = \ g_1 J_1^6 + g_2 J_2^3 + g_3 J_3^2 + g_4 J_1^4 J_2 + g_5 J_1^3 J_3 + g_6 J_1^2 J_2^2 + g_7 J_1 J_2 J_3 \ ; $$
for the latter we have
$$ \begin{array}{rl}
\de \Phi_6 \ =& \
(40 b2  \th_1) J_1^6 +
( 8 b_1 \th_4 ) J_2^3 +
(12 b_1  \th_5 ) J_3^2 \\ & +
(40 b_1 \th_1 + 4 b_1 \th_2 + 40 b_2 \th_2) J_1^4 J_2 \\ & +
(12 b_1 \th_2 + 8 b_1 \th_3 + 40 b_2 \th_3) J_1^3 J_3 \\ & +
(24 b_1 \th_2 + 8 b_1 \th_4 + 40 b_2 \th_4) J_1^2 J_2^2 \\ & +
(16 b_1 \th_3 + 24 b_1 \th_4 + 12 b_1 \th_5 + 40 b_2 \th_5)
J_1 J_2 J_3 \end{array} $$
Thus, with by now standard notation, we get
$$ \^g \ = \ g \ + \ M_g \theta $$
where the matrix $M_g$ is
$$ M_g \ = \ 4 \ \pmatrix{
10 b_2 & 0 & 0 & 0 & 0 \cr
0 & 0 & 0 & 2 b_1 & 0 \cr
0 & 0 & 0 & 0 & 3 b_1 \cr
10 b_1 & (b_1 + 10 b_2) & 0 & 0 & 0 \cr
0 & 3 b_1 & (2 b_1 + 10 b_2) & 0 & 0 \cr
0 & 6 b_1 & 0 & (2 b_1 + 10 b_2) & 0 \cr
0 & 0 & 4 b_1 & 6 b_1 & (3 b_1 + 10 b_2) \cr} $$
We can in general eliminate five out of the seven coefficients; due to the sparse nature of $M_g$, however, we have several limitation. A complete table giving conditions for each allowed choice would be rather long, and is easily built by considering the determinant of the corresponding submatrices; it is thus omitted.

In view of comparison with \cite{SGU}, it is however interesting to consider the choices which do not set $g_3$ to zero; for these to be possible the determinant of the corresponding submatrix must be zero, and hence we have the following table of conditions for each choice to be possible:
$$ \begin{array}{ll}
g_2 = g_4 = g_5 = g_6 = g_7 = 0 & b_1 (b_1 + 5 b_2) (3 b_1 + 10 b_2) \not= 0 \\
g_1 = g_4 = g_5 = g_6 = g_7 = 0 & b_2 (b_1 + 5 b_2) (b_1 + 10 b_2) (3  b_1 + 10 b_2) \not= 0 \\
g_1 = g_2 = g_5 = g_6 = g_7 = 0 & b_1 b_2 (b_1 + 5 b_2) (3 b_1 + 10 b_2) \not= 0 \\
g_1 = g_2 = g_4 = g_6 = g_7 = 0 & {\rm corresponding  \ determinant \ is \ zero} \\
g_1 = g_2 = g_4 = g_5 = g_7 = 0 & b_1 b_2 (b_1 + 5 b_2) (b_1 + 10 b_2) (3 b_1 + 10 b_2) \not= 0 \\
g_1 = g_2 = g_4 = g_5 = g_6 = 0 & {\rm corresponding  \ determinant \ is \ zero.} \end{array} $$
Needless to say, when the corresponding determinant is zero, the choice is never allowed.

\subsection{Comparison with the paper by Sergienko, Gufan and Urazhdin \cite{SGU}}

In \cite{SGU} the SGU model is discussed for a fixed (noncritical) value of the control parameter, identified with $a$ in our notation. It is claimed that one can always set to zero all coefficients at the exception of six; in the present notation these correspond to $a, b_1 , b_2 , c_1 , d_2 , g_3$. The statement of \cite{SGU} is based on the observation that all the other terms lie in the range of the $\P$ matrix.

Here we have discussed the SGU model for $a$ in a range of values including the critical value $a=0$. Hence we required that only terms which lie in the range of the matrix $\P$ {\it when this is applied to vectors which are uniformly different from zero} throughout a range of parameters including those at which the phase transition takes place, are eliminated. It is thus not surprising that we get results which are not coinciding with those of \cite{SGU}.

As mentioned above, the SGU approach corresponds to working at a fixed value of the control parameter, arbitrarily near but however bounded away from the phase transition point; needless to say, this can also provide a wealth of useful informations about the behaviour of the system near the phase transition.

Thus, the present comparison is just aimed at showing to which extent one can reproduce the result of \cite{SGU} by the simple and straightforward method exposed here. On the other hand, it also shows which of the possible extra terms should be checked with the more powerful methods of singularity theory, i.e. for which extra terms in Landau potential one has to check the qualitative behaviour is not affected.

Let us now look in detail at the SGU reduced Landau potential from the point of view of the discussion conducted in the present section. (Recall we assume throughout that $b_1 \not= 0$, $b_2 \not= 0$.)

Summarizing our discussion for the SGU model with varying control parameter, we have shown that if $b_1,b_2$ do not satisfy a set of ``resonance relations'' (detailed above), then we can always reduce to a case where the only nonzero coefficients in the Landau polynomial, beside $a$, $b_1$ and $b_2$, are: one of the $c_i$, one of the $d_1,d_2,d_3$, one of the $f_i$, and two of the $g_i$.

If $b_1,b_2$ satisfy some of the ``resonance relations'' given above, then some of the coefficients can not be set to zero; see the above discussion for the different cases.
Let us look more closely to these from the point of view of comparison with \cite{SGU}, always assuming $b_1$ and $b_2$ are nonzero (this condition will not be mentioned in the following).

We have seen in our discussion of terms of order six that one can set $c_2 = c_3 = 0$ provided $b_1 \not= - 4 b_2$. Note this is an extra condition, not mentioned in \cite{SGU}. On the other hand, according to our discussion other choices would also be possible: in particular, one can always set $c_1 = c_2 = 0$ or $c_1 = c_3 = 0$.

As for order eight terms, we have seen that the choice $d_1 = d_3 = d_4 = 0$ is legitimate provided $b_1$ is neither equal to $3 b_2$ nor to $6 b_2$. Again, on the one hand this restriction is not mentioned in \cite{SGU}, and on the other hand our discussion shows that other choices are also legitimate.

Coming to terms of order ten, all of these are set to zero in \cite{SGU}; our discussion shows that -- as far as the Lie-Poincar\'e theory is concerned -- at least one of these must be kept different from zero if we consider a critical range of values for $a$; If the $b_i$ satisfy some resonance relation, our choice is restricted. To fix ideas, we set $f_2 = f_3 = f_4 = f_5 = 0$, which is fine provided $3 b_1  \not= - 20 b_2$.

Finally, concerning terms of order twelve, in \cite{SGU} these are all set equal to zero except for $g_3$. According to our discussion, and again within the limits of Lie-Poincar\'e theory, one should keep at least two of them different from zero (more if the $b_i$ satisfy resonance relations). It is legitimate to choose $g_3$ as one of these, provided the other is neither $g_5$ nor $g_7$; in other words, one should also keep either one of $g_1, g_2 , g_4 , g_6$ as nonzero.
Again to fix ideas, let us say we choose $g_1$; this is legitimate provided $b_1 \not= - 5 b_2$ and $3 b_1 \not= - 10 b_2$.
\medskip

Summarizing our discussion, in \cite{SGU} a large number of the arbitrary coefficients which in principles would appear in the Landau potential are set to zero on the basis of singularity theory considerations; we have seen that to a large extent this can be justified on the basis of the much simpler theory exposed here. This  makes use only of explicit changes of variables, with coefficients which can be explicitely computed by solving linear algebraic equations alone.

Our simple considerations, on the other hand, would require other two coefficients -- i.e. other two basic monomials, one of order ten and one of order twelve -- to be also present; moreover, on the basis of them one would suspect that the potential analyzed in \cite{SGU} experience some special, i.e. non structurally stable, behaviour (even at orders not higher than eight) when one of the ``resonance conditions'' met in our discussion occur.

It should be stressed again that these differences do {\it not} entail the claim that there is something wrong with the analysis of \cite{SGU}: we have used a specific simple tool to change the coordinate expression of the general potential, while Sergienko, Gufan and Urazhdin allowed for elimination of other terms provided this does not change the qualitative predictions of the model. In other words, in our discussion the potential is not changed (although the coordinate expression changes), while they admit the potential can be  changed to a different one, provided the two are qualitatively equivalent.

\section{Conclusions}
\def\sn{6}

In the Landau theory of phase transitions, one considers an effective potential $\Phi$ whose symmetry group $G$ and degree $d$ depend on the Physics of the system under consideration.

As a rule, one should consider as $\Phi$ the most general $G$-invariant polynomial of degree $d$, the latter being chosen on the basis of thermodinamic stability considerations.
When such a $\Phi$ turns out to be too complicate for a direct analysis, it is essential to be able to drop ``unessential terms'', i.e. to consider a simplified potential $\^\Phi$ giving raise to a behaviour qualitatively equivalent to that generated by the general one.

Criteria based on singularity theory and employing the spectral sequence technique exist and have a rigorous foundation \cite{ToD,San}, but are mathematically sophisticated and often very difficult to apply in practice.

Here we consider a simplifying criterion stated by Gufan \cite{Guf}, see the introduction. We rigorously justified a closely related reduction criterion (see section 3) on the basis of classical Lie-Poincar\'e theory as far as one deals with fixed values of the control parameter(s) in the Landau potential.

When one considers a range of values for the control parameter(s), in particular near a phase transition, the reduction criterion has to be slightly modified, as we discussed in section 4; in particular, in order to eliminate a higher order term, certain matrices must be invertible for the full range of values of the control parameter(s).

It should be stressed that in many cases one is satisfied with analyzing the behaviour of the Landau potential for fixed values (near the transition point) of the control parameter(s); in these cases the ``fixed parameters'' reduction criterion, more closely related to the Gufan simplifying criterion, has to be used.

In other cases one wants to be able to ``follow'' the critical points of the Landau potential as the control parameter(s) is (are) changed; in this case one should use the modified reduction criterion given in section 4.

The theory exposed here does not just provide a rigorous proof of the validity of the reduction criteria: it also allows to make completely explicit computations: one passes from $\Phi$ to $\^\Phi$ by a sequence of changes of variables of a well defined form, and depending on a finite number of constants; the values of the latter can moreover be explicitely computed by solving linear algebraic equations.

Due to these feature, it is also straightforward to explicitly compute nondegeneracy conditions -- ensuring the method can be actually applied to the problem at hand -- in terms of the coefficient of low order terms in the Landau potential $\Phi$: indeed, the nondegeneracy conditions correspond simply to the nonvanishing of the determinants of the matrices which must be inverted in order to implement the algorithm.

We have considered three specific cases in detail. Two of these -- i.e. examples 4 and 5 -- correspond to a two-dimensional order parameter $(x,y)$, with the group $G$ consisting, respectively, of $G = \{ I , R_x , R_y , R_{xy} \} $ in example 4, and of $G = \{ I , R_{xy} \}$ in example 5. Here $I$ is the identity, $R_x$ ($R_y)$ is the reflection in $x$ (in $y$), and $R_{xy}$ the reflection in both $x$ and $y$.

In the third case studied in detail, corresponding to example 6 and the Sergienko-Gufan-Urazhdin model, the order parameter is three-dimensional and the group $G$, with a notation analogous to the one just used, is generated by $\{ R_x , R_y , R_z \}$, i.e. consists of $G = \{ I , R_x , R_y , R_z , R_{xy} , R_{xz} , R_{yz} , R_{xyz} \}$.

In the final section, we studied the Sergienko-Gufan-Urazhdin model in detail; in particular, we pointed out that there are some terms which can be eliminated at a given (noncritical) value of the control parameter(s), but which should be retained -- according to our method -- if one wants to study a full range of values of the parameter(s) including critical ones.

Finally, we note that the algorithmic procedure described here can be easily  performed by means of an algebraic manipulation language (like MAPLE, MATHEMATICA, MATLAB...); this also means that one can effectively tackle the problem of simplifying rather complicated potentials in this way.

\end{document}